\documentclass[11pt,titlepage]{article}

\usepackage[english]{babel}
\usepackage[utf8]{inputenc} 
\usepackage{graphicx,tikz}

\usepackage{latexsym}%
\usepackage{amsfonts, amsmath}%
\usepackage[amssymb]{SIunits}%
\usepackage{authblk}%

\usepackage{multicol}

\usepackage{comment}
\usepackage{adjustbox}
\usepackage{wrapfig}

\usepackage[%
letterpaper,%
top=25mm,%
bottom=25mm,%
inner=25mm,%
outer=25mm
]{geometry}

\usepackage[%
bookmarksnumbered,%
pdfauthor={Escuela de Ciencias Fisicas y Matematicas ECFM-USAC},%
pdftitle={XI SILAFAE GUATEMALA 2016},%
pdfsubject={To present the planning of the activity},%
pdfkeywords={silafae, guatemala}%
]{hyperref}
\usepackage
{xcolor}

\usepackage{amsmath}
\usepackage{amsfonts}
\usepackage{amssymb}
\usepackage{bbm}
\usepackage{cancel} 
\usepackage{youngtab}

\usepackage{graphicx}
\usepackage{feynmp} 
\usepackage{subfigure} 
\usepackage{float} 

\usepackage{multirow} 

\usepackage{pgfplots}
\usepackage{xparse}

\usetikzlibrary{positioning,arrows,patterns}
\usetikzlibrary{decorations.markings}
\usetikzlibrary{calc}

\tikzset{
  photon/.style={decorate, decoration={snake}, draw=black},
  fermion/.style={draw=black, postaction={decorate},decoration={markings,mark=at position .55 with {\arrow{>}}}},
  vertex/.style={draw,shape=circle,fill=black,minimum size=3pt,inner sep=0pt},
  vacuum/.style={draw,shape=rectangle,fill=black,minimum size=3pt,inner sep=0pt},
}

\NewDocumentCommand\semiloop{O{black}mmmO{}O{above}}
{%
\draw[#1] let \p1 = ($(#3)-(#2)$) in (#3) arc (#4:({#4+180}):({0.5*veclen(\x1,\y1)})node[midway, #6] {#5};)
}

\begin{document}

\title{A new U(1) model anomaly free for three families to address fermion mass hierarchy and neutrino physics}

\author[1]{S. F. Mantilla, R. Martinez}
\affil[1]{Universidad Nacional de Colombia, Bogot\'a, Colombia}
\date{}

\maketitle
\abstract{
A new non-universal abelian extension $\mathrm{G_{SM}}\otimes \mathrm{U(1)}_{X}$ to the Standard Model free from chiral anomalies is presented. The new $X$ charges distinguish among generations so as the fermion mass matrices get zero-textures from the fermion mass hierarchy can be obtained. Regarding to neutrino physics, the model inclues Majorana fermions which induce see-saw mechanisms to get small neutrino masses. Moreover, the suitability of the model is tested by searching for regions in the parameter space which reproduce current neutrino oscillation data. 
}

\vspace{1cm}
The Standard Model of particle physics\cite{SM} (SM) comprises one of the most successful frameworks within lots of phenomenological facts can be understood at a high level precission. However, some facts lie out of its scope, suggesting the existence of physics beyond the SM (BSM). One of the most important issues discusses the mass fermion hierarchy, the fact that fermions masses span many orders of magnitude, from units of MeV to hundreds of GeV. Some models have been proposed schemes in order to solve it.

One of the earliest schemes which adressed the fermion hierarchy came from left-right models $\mathrm{\mathrm{SU}(2)}_{L}\otimes \mathrm{\mathrm{SU}(2)}_{R}\otimes \mathrm{\mathrm{U}(1)}_{B-L}$ in which is possible to get Fritzsch zero-texture mass matrices\cite{fritzsch1978}. This structure may be useful in predicting mass hierarchy, mixing angles and CP violation phase, in accordance with phenomenological data. Similarly, the lepton sector can also get such a zero-texture as Fukugita, Tanimoto and Yanagida showed in \cite{fty1993}. Moreover, the smallness of neutrino masses and the matter-antimatter asymmetry are understood due to the existence of Majorana fermions which produce not only see-saw mechanisms\cite{inverseseesaw,catano2012} but also violate lepton number conservation through leptogenesis\cite{leptogenesis}.

Regarding to neutrino physics, the massive nature of them has been proved by neutrino oscillations which are another phenomena not predicted by the SM. Many experiments around the world have observed these phenomena since Homestake\cite{homestake} to NO$\nu $A \cite{nova}. The references \cite{neutrinodata, esteban-gonzalez-maltoni}, available at NuFIT 3.0 \cite{nufit}, summarize the reports and fit the current data to neutrino oscillations model. Two hierarchies are presented: normal ordering (NO) in which the squared mass difference between the third and first species accomplish $\Delta m_{31}^2 > 0$, and inverted ordering (IO) in which $\Delta m_{32}^2 < 0$ between the second and third species.  

A large number of schemes have been proposed to solve these issues, but abelian extensions may be not only the simplest, but also enough to understand them\cite{moretti, zprime-review}. Inside these there are non-universal extensions which address fermion mass hierarchy\cite{somepheno} and dark matter\cite{DM-martinez-I, DM-martinez-II, DM-jhep}. The present model is an improvement of \cite{somepheno} because the leptons are now non-universal so as their mass herarchy can also be addressed in the same way as the quark sector. Furthermore, the suitability of the model is tested by exploring its parameter space in search for regions were neutrino oscillation data\cite{esteban-gonzalez-maltoni,neutrinodata} can be reproduced. 

Following the introduction, the particle content of the model is presented. Thereafter, the fermion mass hierarchy is addressed, in which the mass matrices of quark and lepton sectors are diagonalized in detail. After that, it is shown how the neutral lepton sector parameters are explored in search for regions where neutrino oscillation data can be gotten, finalizing with some conclusions.

\vspace{1cm}

The present model proposes an abelian extension $\mathrm{G_{SM}}\otimes \mathrm{U(1)}_{X}$ whose quantum numbers are non-universal in the fermionic sector and free from chiral anomalies: $\left[\mathrm{\mathrm{SU}(3)}_{C} \right]^{2} \mathrm{\mathrm{U}(1)}_{X}$, $\left[\mathrm{\mathrm{SU}(2)}_{L} \right]^{2} \mathrm{\mathrm{U}(1)}_{X}$, $\left[\mathrm{\mathrm{U}(1)}_{Y} \right]^{2}   \mathrm{\mathrm{U}(1)}_{X}$, $\mathrm{\mathrm{U}(1)}_{Y}   \left[\mathrm{\mathrm{U}(1)}_{X} \right]^{2}$, $\left[\mathrm{\mathrm{U}(1)}_{X} \right]^{3}$ and $\left[\mathrm{Grav} \right]^{2}   \mathrm{\mathrm{U}(1)}_{X}$. 

A discrete symmetry $\mathbb{Z}_{2}$ has been added in order to get at least one generation distinguishable. Such a set of quantum numbers is obtained by including exotic fermions, so as the number of non-universal solutions get increased. These new fields are chiral isospin singlets, so they must get mass through a new scalar singlet named $\chi$ which also breaks the abelian extension through its vacuum expectation value (VEV) $v_{\chi}$. Additionally, a scalar singlet $\sigma$ with the inverse $\mathbb{Z}_{2}$ parity of $\chi$ and without VEV is also added such that radiative corrections involving such a field give mass to massless fermions at tree level. Lastly, since the active neutrinos have really small masses, the particle content has a set of three Majorana fermions so as the inverse see-saw mechanism (ISS) can be implemented. 

The previous requirements are satisfied by the fermion spectrum shown in table \ref{tab:Fermionic-content}. In addition to the SM fermions, there were added three right-handed neutrinos $\nu_{R}^{e,\mu,\tau}$, one up-like quark $T$, two down-like quarks $J^{1,2}$ and three Majorana fermions ${N}_{R}^{1,2,3}$. Regarding to the scalar sector, shown in table \ref{tab:Scalar-content}, two Higgs doublets with different $X$-charges were included together with the scalar singlets $\chi$ and $\sigma$. Finally, the gauge sector comprises the SM gauge bosons and the new $Z_{\mu}'$ associated with $\mathrm{U(1)}_{X}$. 

\begin{table}
\centering
\begin{adjustbox}{width=0.62\textwidth,center}
\begin{tabular}{cccc|cccc|}
\hline\hline
Quarks	&	$X$	&$\mathbb{Z}_{2}$&&	Leptons	&	$X$&$\mathbb{Z}_{2}$	\\ \hline 
\multicolumn{7}{c}{\small{SM Fermionic Isospin Doublets}}	\\ \hline\hline
\small{$q^{1}_{L}=\left(\begin{array}{c}U^{1} \\ D^{1} \end{array}\right)_{L}$}
	&	$+1/3$	&$+$	&&
\small{$\ell^{e}_{L}=\left(\begin{array}{c}\nu^{e} \\ e^{e} \end{array}\right)_{L}$}
	&	$0$	&$+$	\\
\small{$q^{2}_{L}=\left(\begin{array}{c}U^{2} \\ D^{2} \end{array}\right)_{L}$}
	&	$0$	&$-$	&&
\small{$\ell^{\mu}_{L}=\left(\begin{array}{c}\nu^{\mu} \\ e^{\mu} \end{array}\right)_{L}$}
	&	$0$	&$+$		\\
\small{$q^{3}_{L}=\left(\begin{array}{c}U^{3} \\ D^{3} \end{array}\right)_{L}$}
	&	$0$	&$+$	&&
\small{$\ell^{\tau}_{L}=\left(\begin{array}{c}\nu^{\tau} \\ e^{\tau} \end{array}\right)_{L}$}
	&	$-1$	&$+$	\\   \hline\hline

\multicolumn{7}{c}{\small{SM Fermionic Isospin Singlets}}	\\ \hline\hline
\begin{tabular}{c}
	\small{$U_{R}^{1,3}$}\\
	\small{$U_{R}^{2}$}\\
	\small{$D_{R}^{1,2,3}$}
\end{tabular}	&	 
\begin{tabular}{c}
	\small{$+2/3$}\\
	\small{$+2/3$}\\
	\small{$-1/3$}\end{tabular}	&
\begin{tabular}{c}
	\small{$+$}\\
	\small{$-$}\\
	\small{$-$}
\end{tabular}	&&
\begin{tabular}{c}
	\small{$e_{R}^{e,\tau}$}\\
	\small{$e_{R}^{\mu}$}
\end{tabular}	&	
\begin{tabular}{c}
	\small{$-4/3$}\\
	\small{$-1/3$}
\end{tabular}	&	
\begin{tabular}{c}
	\small{$-$}\\
	\small{$-$}
\end{tabular}\\   \hline \hline 

\multicolumn{3}{c}{\small{Non-SM Quarks}}	&&	\multicolumn{3}{c}{\small{Non-SM Leptons}}	\\ \hline \hline
\begin{tabular}{c}
	\small{$T_{L}$}\\
	\small{$T_{R}$}
\end{tabular}	&
\begin{tabular}{c}
	\small{$+1/3$}\\
	\small{$+2/3$}
\end{tabular}	&
\begin{tabular}{c}
	\small{$-$}\\
	\small{$-$}
\end{tabular}	&&
\begin{tabular}{c}
	\small{$\nu_{R}^{e,\mu,\tau}$}\\
	\small{$N_{R}^{e,\mu,\tau}$}
\end{tabular} 	&	
\begin{tabular}{c}
	\small{$1/3$}\\
	\small{$0$}
\end{tabular}	&	
\begin{tabular}{c}
	\small{$-$}\\
	\small{$-$}
\end{tabular}\\
	\small{$J^{1,2}_{L}$}	&	  \small{$0$} 	&\small{$+$}	&&	
	\small{$E_{L},\mathcal{E}_{R}$}	&	\small{$-1$}	&\small{$+$}	\\
	\small{$J^{1,2}_{R}$}	&	 \small{$-1/3$}	&\small{$+$}	&&	
	\small{$\mathcal{E}_{L},E_{R}$}	&	\small{$-2/3$}	&\small{$+$}	\\ 
	\hline \hline
\end{tabular}
\end{adjustbox}
\caption{Non-universal $X$ quantum number and $\mathbb{Z}_{2}$ parity for SM and non-SM fermions.}
\label{tab:Fermionic-content}
\end{table}

\begin{table}
\centering
\begin{adjustbox}{width=0.33\textwidth,center}
\begin{tabular}{ccc}
Scalar bosons	&	$X$	&	$\mathbb{Z}_{2}$	\\ \hline 
\multicolumn{2}{c}{Higgs Doublets}\\ \hline\hline
$\phi_{1}=\left(\begin{array}{c}
\phi_{1}^{+} \\ \frac{h_{1}+v_{1}+i\eta_{1}}{\sqrt{2}}
\end{array}\right)$	&	$2/3$	&	$+$	\\
$\phi_{2}=\left(\begin{array}{c}
 \phi_{2}^{+} \\ \frac{h_{2}+v_{2}+i\eta_{2}}{\sqrt{2}}
\end{array}\right)$	&	$1/3$	&	$-$	\\   \hline\hline
\multicolumn{2}{c}{Higgs Singlets}\\ \hline\hline
$\chi  =\dfrac{\xi_{\chi}  +v_{\chi}  +i\zeta_{\chi}}{\sqrt{2}}$	& $-1/3$	&	$+$	\\   
$\sigma$	& $-1/3$	&	$-$	\\   \hline \hline 
\end{tabular}
\end{adjustbox}
\caption{Non-universal $X$ quantum number for Higgs fields.}
\label{tab:Scalar-content}
\end{table}


The mass acquisition of fermions is produced by the Yukawa Lagrangian of the model. Thus, in order to satisfy the gauge symmetry $\mathrm{SU(2)}_{L}\otimes \mathrm{U(1)}_{Y} \otimes \mathrm{U(1)}_{X}$, some Yukawa couplings are forbidden and the mass matrices get zero textures which produce the fermion mass hierarchy. In the quark sector, the resulting Yukawa Lagrangian is
\begin{eqnarray}
-\mathcal{L}_Q &=& \overline{q_L^{1}}\left(\widetilde{\phi} _2h^{U}_2 \right)_{1j}U_R^{j}+\overline{q_L^{a}}(\widetilde{\phi }_1 h^{U}_{1})_{aj}U_R^{j}+\overline{q_L^{1}}\left(\phi _1 h^{D}_1\right)_{1j}D_R^{j}+\overline{q_L^{a}}\left(\phi _2 h^{D}_{2} \right)_{aj}D_R^{j} \notag \\
&+&\overline{q_L^{1}} (\phi _1 h^{J}_{1})_{1m} J^{m}_R+\overline{q_L^{a}}\left(\phi  _2 h^{J}_{2} \right)_{am} J^{m}_R+\overline{q_L^{1}}\left(\widetilde{\phi} _2 h^{T}_{2} \right)_1T_R +\overline{q_L^{a}} (\widetilde{\phi } _1 h^{T}_{1})_aT_R \notag \\
&+&\overline{T_{L}}\left( \sigma h_{\sigma }^{U}+\chi h_{\chi }^{U}\right)_{j}{U}_{R}^{j}+\overline{T_{L}}\left( \sigma h_{\sigma}^{T}+\chi h_{\chi }^{T}\right){T}_{R}
\nonumber \\
&+&\overline{J_{L}^n}\left( \sigma ^*h_{\sigma }^{D}+\chi ^*h_{\chi }^{D}\right)_{nj}{D}_{R}^{j}+\overline{J_{L}^n}\left( \sigma ^*h_{\sigma }^{J}+\chi ^*h_{\chi }^{J}\right)_{nm}{J}_{R}^{m}+h.c.,
 \label{quark-yukawa-1}
\end{eqnarray}
where $\widetilde{\phi}_{1,2}=i\sigma_2 \phi_{1,2}^*$ are the conjugate scalar doublets, $a=2,3$ label the quark doublets of the 2nd and 3rd families, and $n(m)=1,2$ are the indices of the $J^{n(m)}$ quarks (a sum over the indices $i, a$ and $n$ is understood), while the lepton sector has the neutral lepton Yukawa Lagrangian
\begin{equation}
\begin{split}
-\mathcal{L}_{Y,N} &= 
h_{2e}^{\nu e}\overline{\ell^{e}_{L}}\tilde{\phi}_{2}\nu^{e}_{R} + 
h_{2e}^{\nu \mu}\overline{\ell^{e}_{L}}\tilde{\phi}_{2}\nu^{\mu}_{R} + 
h_{2e}^{\nu \tau}\overline{\ell^{e}_{L}}\tilde{\phi}_{2}\nu^{\tau}_{R} + 
h_{2\mu}^{\nu e}\overline{\ell^{\mu}_{L}}\tilde{\phi}_{2}\nu^{e}_{R} \\ &+
h_{2\mu}^{\nu \mu}\overline{\ell^{\mu}_{L}}\tilde{\phi}_{2}\nu^{\mu}_{R} + 
h_{2\mu}^{\nu \tau}\overline{\ell^{\mu}_{L}}\tilde{\phi}_{2}\nu^{\tau}_{R} +	
h_{\chi i}^{\nu j} \overline{\nu_{R}^{i\;C}} \chi^{*} N_{R} +
\frac{1}{2} \overline{N_{R}^{i\;C}} M^{ij}_{N} N_{R}^{j} + \mathrm{h.c.},
\end{split}
\label{eq:Neutrino-Lagrangian}
\end{equation}
and the charged lepton Yukawa Lagrangian given by
\begin{equation}
\begin{split}
-\mathcal{L}_{Y,E} &= 
\eta \overline{\ell^{e}_{L}}\phi_{2}e^{\mu}_{R} + h \overline{\ell^{\mu}_{L}}\phi_{2}e^{\mu}_{R} + 
\zeta\overline{\ell^{\tau}_{L}}\phi_{2}e^{e}_{R} + H\overline{\ell^{\tau}_{L}}\phi_{2}e^{\tau}_{R} +	
q_{11}\overline{\ell^{e}_{L}}\phi_{1}E_{R} + q_{21}\overline{\ell^{\mu}_{L}}\phi_{1}{E}_{R} \\ &+
h_{\sigma e}^{E}\overline{E_{L}}\sigma e^{e}_{R} + h_{\sigma \mu}^{\mathcal{E}}\overline{\mathcal{E}_{L}}\sigma^{*} e^{\mu}_{R} + 
h_{\sigma \tau}^{E}\overline{E_{L}}\sigma e^{\tau}_{R} + 
H_{1}\overline{E_{L}}\chi E_{R} + H_{2}\overline{\mathcal{E}_{L}}\chi^{*} \mathcal{E}_{R} + \mathrm{h.c.}
\end{split}
\label{eq:Electron-Lagrangian}
\end{equation}
The following subsections presents the mass acquisition in each one of the fermionic sectors, where the implementation of see-saw mechanisms and radiative corrections is extensively used so as every fermion gets mass according to phenomenological data. 

After the SSB, the mass matrix of the up-like quarks is, in the basis $\mathbf{U}=(u^{1},u^{2},u^{3},T)$
\begin{small}
\begin{eqnarray}
\mathbb{M}_{U}=
\frac{1}{\sqrt{2}}\begin{pmatrix}
0 & \upsilon _2 a_{12} & 0 &  \left| \right. & \upsilon _2 y_1   \\
0 & \upsilon _1 a_{22} & 0 &  \left| \right. &  \upsilon _1 y_2 \\
\upsilon _1 a_{31} & 0 & \upsilon _1 a_{33} & \left| \right. &0\\
\text{\textemdash} & \text{\textemdash} & \text{\textemdash} & \text{\textemdash} & \text{\textemdash}\\
0  & \upsilon _{\chi}c_2  & 0   &  \left| \right.  & \upsilon _{\chi}h_{\chi }^T
\end{pmatrix}.
\end{eqnarray}
\end{small}
Its eigenvalues are obtained by diagonalizing its left-handed squared matrix $\mathbb{M}_{U}^{2}=\mathbb{M}_U(\mathbb{M}_U)^\mathrm{T}$.
Since the blocks of $\mathbb{M}_{U}^{2}$ have a hierarchy suited for a see-saw procedure\cite{martinez1612,grimus}, it is implemented so as the exotic species $T$ can be rotated out, obtaining the block-diagonal matrix $\mathbbm{m} _U^2=\mathrm{diag}(m_U^2, m_T^2)$, where $m_T^2$ is the squared mass of the exotic $T$ quark, $m_T^2\approx  \frac{1}{2} \upsilon _{\chi}^2\left(c_2^2+h_{\chi }^{T2}\right)$.

The eigenvalues of the resulting SM up quark mass matrix $m_U^2$ are the masses of the $t$ and $c$ quarks
\begin{eqnarray}
m_{c}^{2}= \upsilon _1^2  \frac{\left(a_{22}h_{\chi }^{T}-y_2c_2\right)^{2}}{2({c_2 ^2+h_{\chi }^{T2}})},\qquad
m_{t}^{2}= \frac{1}{2}\upsilon _1 ^2 \left(a_{31}^2+a_{33}^2 \right),
\label{top-mass}
\end{eqnarray}
However, the $u$ quark mass have to be generated through radiative corrections\cite{martinez-rad-corr-331}. By adding up such contributions to the mass matrix and diagonalizing it again, the mass of the $u$ quark turns out to be
\begin{eqnarray}
m_u^2=\frac{1}{2}\upsilon _1^2 \Sigma _{11}^2,
\label{up-mass}
\end{eqnarray}
where $\Sigma _{11}$ is guven by
\begin{eqnarray}
\Sigma _{11}=\frac{-1}{16\pi ^2}\frac{f'\left(h^U_{\sigma}\right)_1\left(h^T_2\right)_1}{\sqrt{2}M_T}C_0\left(\frac{M_2}{M_T},\frac{M_{\sigma}}{M_T}\right),
\label{oneloop-correction}
\end{eqnarray} 
and the loop coefficient is $C_0\left(x_1,x_2\right)$ is
\begin{eqnarray}
C_0\left(x_1,x_2\right)=\frac{1}{\left(1-x_1^2\right)\left(1-x_2^2\right)\left(x_1^2-x_2^2\right)}\left[x_1^2x_2^2\ln\left(\frac{{x_1^2}}{x_2^2}\right)-x_1^2\ln x_1^2+x_2^2 \ln x_2^2\right].
\label{oneloop-coef}
\end{eqnarray}

In a similar way, the SSB gives the mass matrix of the down-like quarks. In the basis $\mathbf{D}=(d^{1},d^{2},d^{3},J^{1},J^{2})$ it turns out to be
\begin{small}
\begin{eqnarray}
M_{D}=
\frac{1}{\sqrt{2}}\begin{pmatrix}
0 & 0 & 0 &  \left| \right.  & \upsilon _1 j_{11} & \upsilon _1 j_{12}   \\
0 & 0 & 0 &  \left| \right.  & \upsilon _2 j_{21} & \upsilon _2j_{22} \\
\upsilon _2 B_{31} & \upsilon _2 B_{32} & \upsilon _2 B_{33} & \left| \right.  & 0 &0 \\
 \text{\textemdash}  &  \text{\textemdash}  &  \text{\textemdash}  &  \text{\textemdash}  &  \text{\textemdash} &  \text{\textemdash}   \\
0  & 0  & 0  &  \left| \right.  & \upsilon _{\chi }k_{11} &  \upsilon _{\chi }k_{12} \\
0 & 0  &0 &  \left| \right. &  \upsilon _{\chi }k_{21}  &  \upsilon _{\chi }k_{22}
\end{pmatrix}.
\label{mass-matrices-2}
\end{eqnarray} 
\end{small}
The mass eigenvalues are obtained by diagonalizing the left-handed squared matrix
Similarly, the blocks of $\mathbb{M}_{D}^{2}$ have the hierarchy suited for a see-saw procedure\cite{martinez1612,grimus}. Thus, the the masses of the exotic species $J^{1,2}$ are, aproximately
\begin{eqnarray}
m_{J^{1}}^2= \frac{1}{2}\upsilon _{\chi }^2k_{11}^2, \qquad 
m_{J^{2}}^2= \frac{1}{2}\upsilon _{\chi }^2k_{22}^2,
\end{eqnarray}
and the resulting mass matrix $m_D^2$ involving only SM down quark is 
\begin{eqnarray}
m_D^2=\frac{1}{2}\begin{pmatrix}
0 & 0 &  0    \\
0 & 0 & 0  \\
0 & 0 &  \upsilon _2 ^2 \left(B_{31}^2+B_{32}^2+B_{33}^2 \right) 
\end{pmatrix},
\label{SM-down-mass}
\end{eqnarray}
where only the $b$ quark has acquired mass,
\begin{eqnarray}
m_b^2=\frac{1}{2}\upsilon _2 ^2 \left(B_{31}^2+B_{32}^2+B_{33}^2 \right).
\end{eqnarray}
The quarks $d$ and $s$ remain massless, inconsistently with the phenomenological data. Therefore, radiative corrections are needed in the down quark sector, analogously to the issue with the $u$ quark. The new components in the matrix $m_{D}^2$ (\ref{SM-down-mass}) are
\begin{small}
\begin{eqnarray}
&&m_{D(\text{1-loop})}^2= \\
&&\frac{1}{2}\begin{pmatrix}
\upsilon _2^2 \left(\Sigma _{11}^2+\Sigma _{12}^2+\Sigma _{13}^2\right) & \upsilon _1\upsilon _2 \left(\Sigma _{11}\Sigma _{21}+\Sigma _{12}\Sigma _{22}+\Sigma _{13}\Sigma _{23}\right) &  \upsilon _2^2 \left(\Sigma _{11}B_{31}+\Sigma _{12}B_{32}+\Sigma _{13}B_{33}\right)  \\
* & \upsilon _1^2 \left(\Sigma _{21}^2+\Sigma _{22}^2+\Sigma _{23}^2\right)  & \upsilon _1\upsilon _2 \left(\Sigma _{21}B_{31}+\Sigma _{22}B_{32}+\Sigma _{23}B_{33}\right)  \\
* & * &  2m_b^2
\end{pmatrix}, \nonumber 
\label{oneloop-SM-down-mass}
\end{eqnarray}
\end{small}
where $\Sigma_{lj}$ is similar to eq. \eqref{oneloop-correction}. In this way, the masses of the $d$ and $s$ quarks can be expressed as, approximately, 
\begin{eqnarray}
m_d^2\approx \frac{\Sigma_{11}\upsilon _2^2}{2m_b^2}, \qquad 
m_s^2\approx \frac{\Sigma_{22}\upsilon _1^2}{2m_b^2}.
\end{eqnarray} 

By evaluating the neutral lepton Yukawa Lagrangian at the VEVs of the scalar fields, a Majorana mass matrix emerges which, in the basis $\mathbf{N}_{L}=\left(\begin{matrix}{\nu^{e,\mu,\tau}_{L}},\,\left(\nu^{e,\mu,\tau}_{R}\right)^{C},\,\left(N^{e,\mu,\tau}_{R}\right)^{C}\end{matrix}\right)^{\mathrm{T}}$ turns out to be
\begin{equation}
\mathbb{M}_{\nu} = 
\left(\begin{array}{c c c}
0	&	m_{D}^{\mathrm{T}}	&	0	\\
m_{D}	&	0	&	M_{D}^{\mathrm{T}}	\\
0	&	M_{D}	&	M_{M}
\end{array}\right),\qquad \mathrm{where}\qquad  
m_{D} = \frac{v_{2}}{\sqrt{2}}\left(\begin{matrix}
h_{2e}^{\nu e}	&	h_{2e}^{\nu \mu}	&	h_{2e}^{\nu \tau}	\\
h_{2\mu}^{\nu e}&	h_{2\mu}^{\nu \mu}	&	h_{2\mu}^{\nu \tau}	\\
0	&	0	&	0	\end{matrix}\right)	
\label{eq:m_nu_original_parameters}
\end{equation}
is a Dirac mass matrix between $\nu_{L}$ and $\nu_{R}$, $M_{D}=h_{\chi}^{\nu}v_{\chi}/\sqrt{2}$ is a Dirac mass between $\nu_{R}^{c}$ and $N_{R}$ ($h_{N\chi}$ is a $3\times 3$ matrix), and $M_{M}$ is the mass of the Majorana fermions $N_{R}$.

By assuming $M_M \ll m_D$ and $M_D $, the matrix will show $\mathbb{M}_{\nu}$ shows the inverse see-saw mechanism where the smallness of active neutrino masses can be addressed. If the following blocks are defined
\begin{equation}
\begin{split}
\mathcal{M}_{\nu} = \left(\begin{matrix}
m_{D}	\\	0
\end{matrix} \right),\qquad 
\mathcal{M}_{N} = \left(\begin{matrix}
0	&	M_{D}^{\mathrm{T}}	\\
M_{D}	&	M_{M}
\end{matrix} \right),
\end{split}
\end{equation}
and a see-saw diagonalization is performed, the resulting mass matrix is $\mathrm{diag}(m_{\mathrm{light}}	,m_{\mathrm{heavy}})$, where $m_{\mathrm{light}}$ and $m_{\mathrm{heavy}}$ are the Majorana masses of the light and heavy neutral species, respectively. They are given by
\begin{eqnarray}
m_{\mathrm{light}} =	m_{D}^{\mathrm{T}} \left( M_{D} \right)^{-1} M_{M} \left( M_{D}^{\mathrm{T}} \right)^{-1} m_{D}	,\qquad
m_{\mathrm{heavy}} =	\left(\begin{array}{c c}
0	&	M_{D}^{\mathrm{T}}	\\ 
M_{D}	&	M_{M}
\end{array}\right).
\label{eq:Light-neutrino-mass-matrix}
\end{eqnarray}
When the latter mass matrix $m_{\mathrm{heavy}}$ is diagonalized, two quasi-degenerated mass matrices appeared, where the degeneracy is broken by the Majorana mass, $m_{\mathrm{heavy}}' = \frac{M_{M}}{2} \pm M_{D}$.

Finally, the charged lepton sector is presented. First, a remarkable property of this sector is that $\mathcal{E}$ gets decoupled from the other leptons, so it does not contribute in the mass acquisition mechanisms. Therefore, the mass matrix is expressed in the basis $\mathbf{E}=(e^{e},e^{\mu} , e^{\tau}, E)$
\begin{equation}
\mathbb{M}_{E} = \frac{v_{2}}{\sqrt{2}}
\begin{pmatrix}
 0  & \eta & 0 & | & q_{11} t_{\beta} \\
 0 & h  & 0 & | & q_{21} t_{\beta} \\
 \zeta        & 0 & H & | & 0 \\
 - & - & - & -& - \\
 0 & 0 & 0 & | &H_{1} v_{\chi}/v_{2}
\end{pmatrix},
\end{equation}
whose determinant is null, i.e., $e$ remains massless. Again, this issue is overpassed by including one-loop corrections so as the complete mass matrix is
\begin{eqnarray}
\mathbb{M}_{E(\text{1})}=\mathbb{M}_{E}+\Delta \mathbb{M}_{E},
\end{eqnarray}
with $\Delta \mathbb{M}_{E}$ being the radiative corrections
\begin{eqnarray}
\Delta \mathbb{M}_{E}=
\frac{\upsilon _2 }{2}\begin{pmatrix}
\Sigma _{11} & 0 &  \Sigma _{13} &  \left| \right. & 0   \\
\Sigma _{21} & 0 & \Sigma _{23} &  \left| \right. & 0 \\
0 & 0 & 0 & \left| \right. &0\\
\text{\textemdash} & \text{\textemdash} & \text{\textemdash} & \text{\textemdash} & \text{\textemdash}\\
0  & 0  & 0   &  \left| \right. & 0
\end{pmatrix}.
\label{oneloop-shift-lept}
\end{eqnarray} The mass eigevalues are obtained by diagonalizing the left-handed squared matrix $\mathbb{M}_{E}\mathbb{M}_{E}^{\dagger}$, which yields
\begin{equation}
\begin{split}
m_{e}^{2} &= \frac{h^2 \Sigma_{11}^2 v_{2}^2}{2 \left(\eta ^2+h^2\right)}\approx \frac{ v_{2}^2}{2}\Sigma_{11}^2,	\qquad 
m_{\tau}^{2} = \frac{v_{2}^{2}}{2} \left(\zeta^2 + H^2\right)\approx \frac{v_{2}^2}{2}H^2,	\\
m_{\mu}^{2} &= \frac{v_{2}^{2}}{2} \left(\eta^2 + h^2\right)\approx \frac{v_{2}^{2}}{2} h^2,	\qquad 
m_{E}^{2} = \frac{H_{1}^2 v_{\chi}^2}{2}.
\end{split}
\end{equation}
The squared mass matrix $\mathbb{M}_{E}\mathbb{M}_{E}^{\dagger}$ can be diagonalized by blocks\cite{grimus} so as the exotic species $E$ is rotated out. Then, the resulting SM charged lepton mass matrix has rotation matrix
\begin{equation}
\label{eq:electron-muon-rotation}\
{V}_{\mathrm{SM},L}^{E} = 
\begin{pmatrix}
 c_{\alpha_{e\mu}}	&	s_{\alpha_{e\mu}}	&	\frac{\Sigma_{13}}{H}	\\
-s_{\alpha_{e\mu}}	&	c_{\alpha_{e\mu}}	&	\frac{\Sigma_{23}}{H}	\\
-\frac{\Sigma_{13}}{H}	&	-\frac{\Sigma_{23}}{H}	&	1
\end{pmatrix},
\end{equation}
where the angle $t_{\alpha_{e\mu}} = \tan \alpha_{e\mu} \approx \eta/h$ is left as a free parameter in order to fit the model to current neutrino phenomenological data.


The suitability of the model in addresing neutrino physics can be tested by exploring its parameter space and searching for regions were neutrino oscillation data are obtained. For simplicity, $M_{D}$ is assumed diagonal and $M_{M}$ proportional to the identity
\begin{equation}
M_{D} = \left( \begin{matrix}
h_{N\chi 1}	&	0	&	0	\\	0	&	h_{N\chi 2}	&	0	\\	0	&	0	&	h_{\chi N 3}
\end{matrix} \right)\frac{v_{\chi}}{\sqrt{2}}, \qquad 
M_{M} = \mu_{N} \mathbb{I}_{3\times 3}.
\end{equation}
Therefore, by replacing \eqref{eq:m_nu_original_parameters} in \eqref{eq:Light-neutrino-mass-matrix}, the Majorana mass for light neutrinos is gotten
\begin{equation}
\label{eq:Neutrino-mass-matrix}
m_{\mathrm{light}} = \frac{\mu_{N} v_{2}^{2}}{{h_{N\chi 1}}^{2}v_{\chi}^{2}}
\left( \begin{matrix}
	\left( h_{2e}^{\nu e}\right)^{2} + \left( h_{2\mu}^{\nu e} \right)^{2} \rho^{2} 	&
	{h_{2e}^{\nu e}}\,{h_{2e}^{\nu \mu}} + {h_{2\mu}^{\nu e}}\,{h_{2\mu}^{\nu \mu}}\rho^2 	&
	{h_{2e}^{\nu e}}\,{h_{2e}^{\nu \tau}}+ {h_{2\mu}^{\nu e}}\,{h_{2\mu}^{\nu \tau}}\rho^2 	\\
	{h_{2e}^{\nu e}}\,{h_{2e}^{\nu \mu}} + {h_{2\mu}^{\nu e}}\,{h_{2\mu}^{\nu \mu}}\rho^2	&	
	\left( h_{2e}^{\nu \mu} \right)^{2} + \left( h_{2\mu}^{\nu\mu} \right)^{2} \rho^{2}	&	
	{h_{2e}^{\nu \mu}}\,{h_{2e}^{\nu \tau}}+ {h_{2\mu}^{\nu \mu}}\,{h_{2\mu}^{\nu \tau}}\rho^2	\\
	{h_{2e}^{\nu e}}  \,{h_{2e}^{\nu \tau}}+ {h_{2\mu}^{\nu e}}  \,{h_{2\mu}^{\nu \tau}}\rho^2	&	
	{h_{2e}^{\nu \mu}}\,{h_{2e}^{\nu \tau}}+ {h_{2\mu}^{\nu \mu}}\,{h_{2\mu}^{\nu \tau}}\rho^2	&	
	\left( h_{2e}^{\nu \tau} \right)^{2} + \left( h_{2\mu}^{\nu \tau} \right)^{2} \rho^{2}
\end{matrix} \right),
\end{equation}
with $\rho={h_{N\chi 1}}/{h_{N\chi 2}}$. The determinant of this matrix is null, so one neutrino is massless and the squared mass differences determine the masses of the other two. $m_{\mathrm{light}}$ is diagonalized by
\begin{equation}
U_{\mathrm{\nu}}^{\mathrm{T}}\, m_{\mathrm{light}}\, U_{\nu} = m_{\mathrm{light}}^{\mathrm{diag}},
\end{equation}
where the matrix $U_{\mathrm{\nu}}$ transforms the weak eigenstates $\nu_{L}^{e,\mu,\tau}$ into mass eigenstates $\nu_{L}^{1,2,3}$. Consequently, the PMNS matrix is obtained as the product of $U_{\mathrm{\nu}}$ with the rotation matrix of the charged sector ${V}_{\mathrm{SM},L}^{E}$
\begin{equation}
U_{\mathrm{PMNS}} = \left( {V}_{\mathrm{SM},L}^{E} \right)^{\dagger} U_{\nu}.
\end{equation}
Finally, as usual, the PMNS matrix will be parametrized by the CKM angless\cite{Beringer2012}.

The search is done by using MonteCarlo procedures on the Yukawa parameters $h_{2e}^{\nu e}$, $h_{2e}^{\nu \mu}$, $h_{2e}^{\nu \tau}$, $h_{2\mu}^{\nu e}$, $h_{2\mu}^{\nu \mu}$, $h_{2\mu}^{\nu \tau}$ and the angle $\alpha_{e\mu}$. The other two angles of ${V}_{\mathrm{SM},L}^{E}$ described by $\Sigma_{13}/H$ and $\Sigma_{23}/H$ were approximated to $m_{e}/m_{\tau}$, while $h_{2e}^{\nu \mu}$ was set zero to simplify the search. The appropriate mass scale and mass ordering can be obtained by adjusting the outer factor of the mass matrix and the ratio $\rho$. The NO is obtained when ${h_{N\chi 1}}^{2} = 0.02$ and ${\rho}^{2} = 0.5$, while the IO corresponds to ${h_{N\chi 1}}^{2} = 0.025$ and ${\rho}^{2} = 0.625$.	In the same way, the mass scale is set by $v_{2} = 7\mathrm{\,GeV}$, $v_{\chi} = 7\mathrm{\,TeV}$ and $\mu_{N} = 1\mathrm{\,keV}$. These values fix the outer factor of the mass matrix \eqref{eq:Neutrino-mass-matrix} at $50\mathrm{\,meV}$, yielding the suited squared-mass differences for NO and IO scenarios\cite{martinez1612}. 

\begin{table}
\centering
\begin{tabular}{c c c c c }
\hline\hline
			&	$\alpha_{e\mu}=0^{\mathrm{o}}$
			&	$\alpha_{e\mu}=15^{\mathrm{o}}$
			&	$\alpha_{e\mu}=30^{\mathrm{o}}$	\\\hline\hline
    $h_{2e}^{\nu e}$	&	$0.264 \rightarrow 0.278$
			&	$0.285 \rightarrow 0.299$
			&	$0.237 \rightarrow 0.270$	\\\hline
    $h_{2e}^{\nu \mu}$	&	$-0.707 \rightarrow -0.244$
			&	$-0.726 \rightarrow -0.335$
			&	$-0.796 \rightarrow -0.547$	\\
    $h_{2\mu}^{\nu \mu}$&	$-0.491 \rightarrow -0.190$
			&	$-0,464 \rightarrow -0.173$
			&	$-0.342 \rightarrow -0.039$	\\\hline
    $h_{2e}^{\nu \tau}$	&	$0.267 \rightarrow 0.748$
			&	$0.313 \rightarrow 0.677$
			&	$0.140 \rightarrow 0.355$	\\
    $h_{2\mu}^{\nu \tau}$&	$0.130 \rightarrow 0.462$
			&	$0.196 \rightarrow 0.460$
			&	$0.440 \rightarrow 0.510$	\\\hline\hline
\end{tabular}
\caption{Yukawa coupling domain which fulfil at $1\sigma$ neutrino oscillation data for NO reported by \cite{neutrinodata}. $h_{2\mu}^{\nu e}=0$ for simplifying the MonteCarlo search.}
\label{tab:Neutrino-parameters-NO}
\end{table}

\begin{table}
\centering
\begin{tabular}{c c c c c }
\hline\hline
			&	$\alpha_{e\mu}=0^{\mathrm{o}}$
			&	$\alpha_{e\mu}=1^{\mathrm{o}}$
			&	$\alpha_{e\mu}=2^{\mathrm{o}}$	\\\hline\hline
    $h_{2e}^{\nu e}$	&	$1.094 \rightarrow 1.107$
			&	$1.091 \rightarrow 1.105$
			&	$1.090 \rightarrow 1.103$	\\\hline
    $h_{2e}^{\nu \mu}$	&	$-0.122 \rightarrow -0.106$
			&	$-0.127 \rightarrow -0.113$
			&	$-0.128 \rightarrow -0.118$	\\
    $h_{2\mu}^{\nu \mu}$&	$0.970 \rightarrow 1.060$
			&	$0.980 \rightarrow 1.070$
			&	$1.010 \rightarrow 1.080$	\\\hline
    $h_{2e}^{\nu \tau}$	&	$0.110 \rightarrow 0.127$
			&	$0.122 \rightarrow 0.138$
			&	$0.135 \rightarrow 0.149$	\\
    $h_{2\mu}^{\nu \tau}$&	$0.930 \rightarrow 1.030$
			&	$0.920 \rightarrow 1.010$
			&	$0.910 \rightarrow 0.980$	\\\hline\hline
\end{tabular}
\caption{Yukawa coupling domain which fulfil at $1\sigma$ neutrino oscillation data for IO reported by \cite{neutrinodata}. $h_{2\mu}^{\nu e}=0$ for simplifying the MonteCarlo search.}
\label{tab:Neutrino-parameters-IO}
\end{table}

The results of the search are presented in tables \ref{tab:Neutrino-parameters-NO} and \ref{tab:Neutrino-parameters-IO} for NO and IO schemes, respectively. The tables show how the region in the parameter space changes with the value of $\alpha_{e\mu}$ in both schemes, and the larger regions in the NO makes the model more consistent with this scheme than the other, according with neutrino oscillation data reported by \cite{neutrinodata} at $3\sigma$. Thus, the neutral spectrum of the model is composed by three active light neutrinos $\nu^{1,2,3}_{L}$ and six quasidegenerated steril neutrinos $N^{1,2,3}_{1L}$ and $N^{1,2,3}_{2L}$ at the TeV scale. 

\vspace{1cm}
Some issues which the SM cannot address, such as the fermion mass hierarchy or neutrino oscillations might be understood in implementing non-universal abelian extensions. This work shows how these schemes are useful in searching for frameworks where fermion masses get their particular ordering without unpleasant fine-tunings. 

From the cancelation of chiral anomalies, the addition of exotic fermions and a supplementary discrete symmetry $\mathbb{Z}_{2}$ the fermion families can be distinguishable so as the mass matrices get zero-textures. Although the mass hierarchy is obtained, too many fermions turned out to be massless and they need radiative corrections to acquire mass. These processes are mediated by the exotic fermions and the scalar field $\sigma$ in such a way that all charged fermions got masses at one-loop level. 
 
On the other hand, the smallness of neutrino masses is achieved by the addition of Majorana fermions which produce an inverse see-saw mechanism and violating lepton number conservation. Because of the null determinant of the neutrino mass matrix, one neutrino gets massless. Additionally, in order to test the suitability of the model in understanding neutrino oscillation data, the parameter space of neutrino Yukawa Lagrangian is explored. The model results more consistent with NO scheme than IO ones. 

\subsubsection*{Acknowledgment}
This work was supported by COLCIENCIAS in Colombia.


\begin{thebibliography}{100}

\bibitem{SM} S.L. Glashow, Nucl. Phys. 22, 579 (1961); S. Weinberg, Phys. Rev. Lett. 19,
1264 (1967); A. Salam, in \textit{Elementary Particle Theory: Relativistic Groups and
Analyticity (Nobel Symposium No. 8)}, edited by N.Svartholm (Almqvist and
Wiksell, Stockholm, 1968), p. 367.

\bibitem{textures} H. Fritzsch, Phys. Lett. B70, 436 (1977) ;  B73, 317 (1978); Nucl. Phys. B155, 189 (1979); H. Fritzsch and Z. Z. Xing, Phys. Lett. 555, 63 (2003); A. Carcamo, R. Martinez and J.-A. Rodriguez, Eur. Phys. J. C50, 935 (2007).  

\bibitem{fritzsch1978} Fritzsch, Harald. Weak-interaction mixing in the six-quark theory. Phys. Lett. B73, 317 (1978).

\bibitem{fty1993} Fukugita, Masataka, Morimitsu Tanimoto, and Tsutomu Yanagida.  Progress of Theoretical Physics 89, 263 (1993).

\bibitem{inverseseesaw} R.  N.	Mohapatra,	Phys. Rev. Lett. 56 (1986), 561; R. N. Mohapatra, J. W. F. Valle, Phys. Rev. D34, 1642 (1986).

\bibitem{catano2012} E. Catano, R. Martinez, and F. Ochoa. Phys.Rev. D86, 073015; A. G. Dias, C. A. de S. Pires, P. S. Rodrigues da Silva, and A. Sampieri. Phys. Rev. D 86, 035007.

\bibitem{leptogenesis} Fukugita, M., and Tsutomu Yanagida.  Phys. Lett. B174, 45 (1986).

\bibitem{homestake} Davis Jr, R., Harmer, D. S., Hoffman, K. C.  Phys. Rev. Lett. 20, 1205 (1968).

\bibitem{nova} 
  J. Bian [NO$\nu$A Collaboration],
  ICHEP2016, Chicago, 06 August, 2016; 	
  P.~Adamson {\it et al.} [NO$\nu$A Collaboration],
  Phys.\ Rev.\ Lett. 116 (2016) no.15,  151806.

\bibitem{neutrinodata} Gonzalez-Garcia, M. C., Michele Maltoni, and Thomas Schwetz. Nucl. Phys. B908 (2016): 199-217.

\bibitem{fritzsch2013} Fritzsch, Harald, and Shun Zhou. Phys. Lett. B718.4: 1457-1464 (2013).

\bibitem{esteban-gonzalez-maltoni} Esteban, I. {\it et al.} arXiv:1609.01864v1 [hep-ph] (2016).


\bibitem{nufit} \url{www.nu-fit.org}.

\bibitem{moretti} L. Basso, S. Moretti, G. Marco Pruna JHEP 1108, 122 (2011); E. Accomando, L. Delle Rose, S. Moretti, E. Olaiya, C. H. Shepherd-Themistocleous, JHEP 1704, 081 (2017);  

\bibitem{zprime-review} A. Leike, Phys. Rep. 317, 143 (1999); J. Erler, P. Langacker, and T. J. Li, Phys. Rev. D 66, 015002 (2002); S. Hesselbach, F. Franke, and H. Fraas, Eur. Phys. J. C 23, 149 (2002).

\bibitem{somepheno} R. Martínez, J. Nisperuza, F. Ochoa, and J. P. Rubio. Phys. Rev. D 89, 056008.

\bibitem{DM-martinez-I} Martinez, R., Nisperuza, J., Ochoa, F., and Rubio, J. P.  Scalar dark matter with CERN-LEP data and $Z'$  search at the LHC in an $\mathrm{\mathrm{U}(1)}'$ model. Physical Review D 90.9 (2014): 095004.

\bibitem{DM-martinez-II} Martinez, R., Nisperuza, J., Ochoa, F., Rubio, J. P., and Sierra, C. F. Phys. Rev. D92.3 (2015): 035016.

\bibitem{DM-jhep} Martinez, R., Ochoa, F., JHEP 05, 113 (2016).

\bibitem{martinez1612} S. F. Mantilla, R. Martinez, and F. Ochoa. Phys. Rev. D 95 (2016): 095037.

\bibitem{grimus} W. Grimus and L. Lavoura, J. High Energy Phys. 11, 042 (2000).

\bibitem{martinez-rad-corr-331} Cárcamo A.E., Martinez, R., and Ochoa, F. Physical Review D 87.7 (2013): 075009.

\bibitem{Beringer2012} Beringer, Juerg, {\it et al.} Phys. Rev. D86.1 (2012): 010001.

\end{thebibliography}
\end{document}